\pdfoutput=1

\documentclass[particles,article,accept,moreauthors,pdftex,10pt,a4paper]{Definitions/mdpi} 
\usepackage{color,soul}

\firstpage{1} 
\pubvolume{0}
\issuenum{1}
\articlenumber{0}
\pubyear{2019}
\copyrightyear{2019}
\history{Received: 30 November 2018; Accepted: 9 January 2019; Published: date}

\updates{yes} % If there is an update available, un-comment this line

\Title{Causality and Renormalization in Finite-Time-Path Out-of-Equilibrium $\phi^3$ QFT}

\Author{Ivan Dadi\'c $^{1}$ and Dubravko Klabu\v{c}ar $^{2,*}$\orcidA{}}

\AuthorNames{Ivan Dadi\'c and Dubravko Klabu\v{c}ar}

\address{%
$^{1}$ \quad Rudjer Bo\v skovi\'c Institute, P.O. Box 180, 10002 Zagreb, Croatia; Ivan.Dadic@irb.hr\\
$^{2}$ \quad Physics Department, Faculty of Science-PMF, University of Zagreb, Bijeni\v{c}ka c. 32, 
10000 Zagreb, Croatia}
\corres{Correspondence: klabucar@phy.hr; Tel.: +385-91-5866730}

\abstract{Our aim is to contribute to quantum field theory (QFT) formalisms
useful for descriptions of short time phenomena, dominant especially in
heavy ion collisions. We formulate out-of-equilibrium QFT
within the finite-time-path formalism (FTP) and renormalization theory (RT).
The potential conflict of FTP and RT is investigated in $g \phi^3$ QFT, 
by using the retarded/advanced ($R/A$) basis of Green functions and
 dimensional renormalization (DR).
For example, vertices immediately after (in time) divergent self-energy loops
do not conserve energy, as integrals diverge. We ``repair'' them, while keeping
 $d<4$, to obtain energy conservation at those vertices.
Already in the S-matrix theory, the renormalized, finite part of Feynman
self-energy $\Sigma_{F}(p_0)$ does not vanish when $|p_0|\rightarrow\infty$
 and cannot be split to retarded and advanced part{s}.
In the Glaser--Epstein approach, the causality is repaired in the composite
 object $G_F(p_0)\Sigma_{F}(p_0)$.
 In the FTP approach, after repairing the vertices, the corresponding composite
 objects are $G_R(p_0)\Sigma_{R}(p_0)$ and $\Sigma_{A}(p_0)G_A(p_0)$.
 In the limit $d\rightarrow 4$, one obtains causal QFT.
The tadpole contribution splits into diverging and finite parts. The 
 diverging, constant component is eliminated by the renormalization
 condition $\langle 0|\phi|0\rangle =0$ of the S-matrix theory.
The finite, oscillating energy-nonconserving tadpole contributions
 vanish in the limit $t\rightarrow \infty $.
}


\begin{document}

%%%%%%%%%%%%%%%%%%%%%%%%%%%%%%%%%%%%%%%%%%

%% Only for the journal Gels: Please place the Experimental Section after the Conclusions

%%%%%%%%%%%%%%%%%%%%%%%%%%%%%%%%%%%%%%%%%%

\section{Introduction and Survey}

In many regions of physics, the {interacting} processes
are embedded in {a} medium and require {a}
 short-time description. To respond to such demands{,}
 neither vacuum S-matrix field theory
 \cite{Giambagi:1972,Thooft:1972,Ashmore:1972,Cicuta:1972,Wilson:1973},
{nor} equilibrium QFT \cite{Donoghue:1983,erratumDonoghue:1983,
Chapman:1997,Nakkagawa:1997,Baacke:1998,Esposito:1998,Knoll:2002,
Jakovac:2005,Arrizabalaga:2007,Blaizot:2007,Blaizot:2015} with the
Keldysh-time-path \cite{Schwinger:1961,Keldysh:1964,KadanoffBook,
Danielewicz:1984,Rammer:1986,Landsman:1987, Calzetta:1988,LeBellacBook,Brown:1998,
 Blaizot:2002,Dadic:2001bp,erratumDadic:2001bp} suffice. The features, a short time
 after the beginning of evolution, where uncertainty relations do not keep energy
 conserved, are to be treated with the finite-time-path {method}. 
Such an approach includes many specific features that are not yet completely understood.
 A particular problem, almost untreated, is
 handling of UV divergences of the QFT as seen at finite time. 
 The present paper is devoted to this problem. We consider it in the
 simplest form of $\lambda\phi^3$ QFT, but many of the discussed features will find
their analogs in more advanced QED and QCD.

Starting with perturbation expansion in the coordinate space, one performs the Wigner transform
 and uses the Wick theorem. The propagators, originally appearing in matrix representation,
 are linearly connected to the Keldysh base with {\it R, A}, and {\it K} %should be italics
 components. For a finite-time-path, the lowest order propagators and
  one-loop self-energies taken at
  $t=\infty$ correspond to Keldysh-time-path propagators and one-loop
  self-energies. For simplicity,
  the label ``$\infty $'' is systematically omitted throughout the paper,
 {except in the Appendix with technical details}.

To analyze the vertices, one further separates $K$-component
 \cite{Dadic:2001bp,erratumDadic:2001bp}
 into its retarded ({\it K,R}) and advanced ({\it K,A}) part{s:}

\begin{eqnarray}\label{RAK}
&&G_{R}(p)=G_{A}(-p)=
{-i \over p^2-m^2+2ip_0\epsilon},
\cr\nonumber\\
&&G_{K}(p{)}=
2\pi\delta(p^2-m^2)[1+2f(\omega_p)]
\cr\nonumber\\
&&
=G_{K,R}(p)-G_{K,A}(p),
\cr\nonumber\\&&
G_{K,R}(p)=-G_{K,A}(-p)=
h(p_0,\omega_p)G_{R}(p),
\cr\nonumber\\&&
\omega_p={\sqrt{\vec p^2+m^2}},~~
\, h(p_0,\omega_p)= \, -\, {p_0 \over \omega_p}\, [1+2f(\omega_p)].
\end{eqnarray}

Matrix propagators are
($i~$ and $~j$ take the values $1,2$):

\begin{eqnarray}\label{RAKM}
&&G_{ij}(p)
={1\over 2}[G_{K}(p{)}+(-1)^jG_{R}(p)+(-1)^iG_{A}(p)].
\end{eqnarray}

Specifically:

\begin{eqnarray}\label{GF}
G_{F}(p)=G_{11}(p)_{f(\omega_p)=0}={-i\over p^2-m^2+2i\epsilon}
,~~G_{\bar F}(p)=-G^*_{F}(p).
\end{eqnarray}

%

%%%%%%%%%%%%%%%%%%%%%%%%%%%%%%%%%%%%%%%%%%

\section{Results}

\vspace{-6pt}
%%%%%%%%%%%%%%%%%%%%%%%%%%%%%%%%%%%%%%%%%%

\subsection{Conservation and Non-Conservation of Energy at Vertices}

Having done all this, one obtains the vertex function (for simplicity,
all the four-momenta are arranged to be incoming to the vertex).
 For the simplicity of discussion, all the times corresponding to the external
 vertices {($j$)} of the whole diagram are assumed equal
 {(}$x_{0,j,ext}=t$, all $~j$; otherwise, some factors,
oscillating with time, but inessential for our discussion, 
would appear{), so that the vertex function~becomes:}
\begin{eqnarray}\label{xsimp}
{i\over 2\pi}{e^{-it\sum_i p_{0i}}
\over \sum_{i}p_{0i}+i\epsilon}.
\end{eqnarray}

This expression \cite{Dadic:2001bp,erratumDadic:2001bp,Dadic:2002}
 integrated over some $dp_{o,k}$ by closing the time-path from
 below gives the expected energy conserving
 $\delta ( \sum_{i}p_{0i})$, with the
 oscillating factor reduced to one. If the integration path catches additional
 singularity, say the propagator's $D(p_k)$ pole at $\bar p_{{0}k}$,
 for this contribution, conservation of energy is ``spoiled'' by a finite amount
$\Delta E=\sum'_i p_{0i}+\bar p_{{0}k}$, and there is {an}
oscillating vertex function
 ${(i/2\pi)} \, {e^{-it\Delta E}/( \Delta E+i\epsilon)}$.
 Note{:} the fact that some time is lower or higher
{than} another,
{ i.e.}, $t_1>t_2$ or $t_1<t_2$, survives Wigner transform
 in the character of ordering (retarded or advanced) of the two-point~function.

 In general, we have {the following possibilities}:

\begin{itemize}[leftmargin=*,labelsep=5.8mm]

\item If the vertex time is lower than the other times of all incoming propagators,
 there are additional contributions, and energy is not conserved at this vertex.
 The oscillations are just what we would expect
 from the Heisenberg uncertainty relations. It is how the time dependence
 emerges in the finite-time-path out{-}of{-}equilibrium QFT. 
 %ADD
 {The ill-defined pinching singularities---products of retarded and
 advanced propagators with the same $(p_0,\vec p)$, only partially eliminated for
 the Keldysh time-path \cite{bedaque:2015}---do not appear here as the propagator energies
 $p_0$ and $p'_0$ are different variables, so that the singularities do not coincide
 except at the point $p_0 = p'_0$. Thus, the pertinent mathematical expressions
 are well defined.}
 %ENDADD

\item For some vertices, at least one incoming propagator $G(p_{{0}k})$
 is advanced (or more generally{, time is lower at the other vertex 
 of this propagator}){; then,} integration over the $p_{0k}$
 (supposed to be UV finite) re{-}establishes energy conservation.

\item The case of UV divergent integrals is interesting; looking {at} 
integrations done separately, one would expect energy conservation, but performing
other integrals before, one notices that the result is ill-defined. The solution is in
regularization: regulated quantities are finite, and (say, in the dimensional regularization)
 the energy conservation is re{-}established (as far as $d<4$).
\end{itemize}

In the $\lambda\phi^3$ QFT, there are two divergent subdiagrams: the tadpole diagram and 
self-energy diagram{, considered separately in the following subsections}.

\subsection{UV Divergence at the Tadpole Subdiagram}

\begin{figure}[H]

\centering

\includegraphics[width=5 cm]{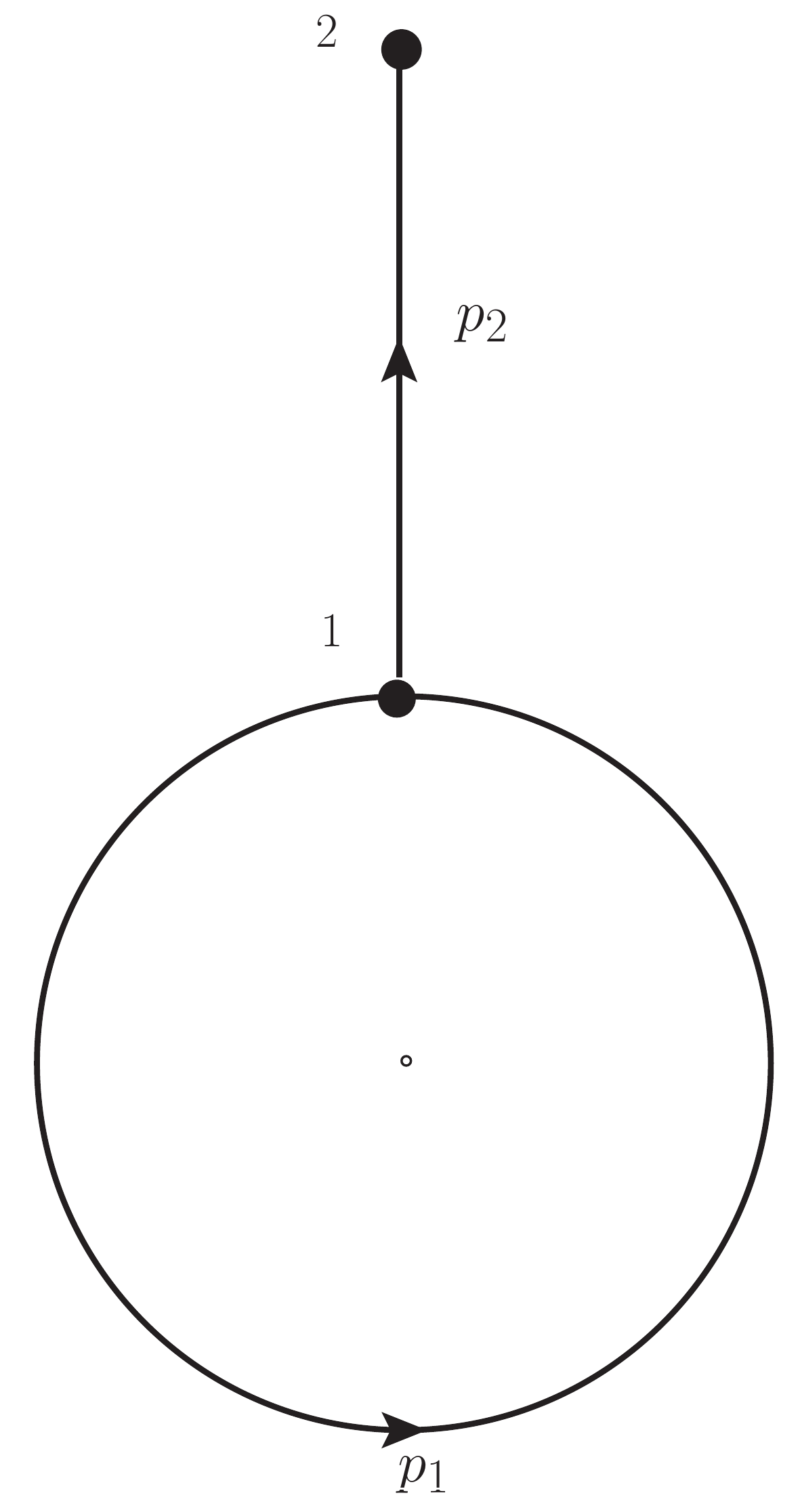}

\caption{The tadpole diagram with a leg.\label{fig1}}

\end{figure}

In the perturbation expansion{, the} tadpole diagram (Figure \ref{fig1}) appears
 as {a} propagator with both ends attached to the same vertex, which is
 {the} (lower-time) end-point vertex of the second propagator.

{The} tadpole subdiagram without {a} leg is simple.
 Of the three components, the loop integral vanishes for the $R$ and $A$
component{s} and diverges for the $K,R$ and $K,A$ {ones}.
 At finite ${\kappa=}\, 4-d$, these integrals are real
 constants related to the $F$ and $\bar F$ components. In the limit $d=4$,
 the renormalization performed on $F$ and $\bar F$ makes them finite.

\begin{eqnarray}\label{gRAK}
&&
ig\mu^{\kappa/2}{\int d^{d}p\over (2\pi)^d}G_{{R}}(p)
=
%\cr\nonumber\\&&
ig\mu^{\kappa/2}{\int d^{d}p\over (2\pi)^d}G_A(p)=0,
\cr\nonumber\\&&
{ G_{Tad} \equiv } - ig{\int d^4p\over (2\pi)^4}G_{K,A}(p) =
- ig\mu^{\kappa/2}{\int d^dp\over (2\pi)^d}{p_0\over \omega_p}{1+2f(\omega_p)\over p^2-m^2-2ip_0\epsilon}
= ig\mu^{\kappa/2}{\int d^dp\over (2\pi)^d}G_{K,R}(p),
\cr\nonumber\\&&
\Longrightarrow -{1\over 2}G_{Tad}=-{igm^2\over 8\pi^2\kappa}-{igm^2\over 16\pi^2}
[1-\gamma_{_E} + \ln({4\pi\mu^2\over m^2})]+{\cal O}(\kappa)
+ig{\int d^3p\over (2\pi)^3}2f(\omega_p)
\cr\nonumber\\&&
=-{igm^2\over 8\pi^2(\kappa)}+finite~vacuum~term +finite~f(\omega_p)~term.
\end{eqnarray}
(Above, and throughout the paper, $\gamma_{_E}$ denotes the Euler-Mascheroni constant,
   $\gamma_{_E} \approx 0.5772$.)

 For a tadpole subdiagram with a leg (see Figure \ref{fig1}), we have
 two vertices; higher in time ($t_2$), which is the connection to the rest of the diagram, and
 lower in time ($t_1,~t_1<t_2$) with the tadpole loop. The lower vertex
 does not conserve energy.

 One has to add
contributions from vertices of Type $1$ and Type $2$.
 We write it symbolically
with the help of the Wigner transform, the connection between the Keldysh-time-path propagators and the
finite-time-path propagators at the time $t'=\infty $ and transition to the $R/A$ basis.
The derivation given in the Appendix shows that:

\begin{eqnarray}\label{gtad}
&&
G_{tad,j}(x_2)=-G_{A}(0,0)
G_{Tad}
%\cr\nonumber\\&&
+\int{dp_{02}\over 2\pi}{ie^{ip_{02}x_{02}}\over p_{02}-i\epsilon}
[G_{A}(p_{02},0)-G_{A}(0,0)]G_{Tad}~.
\end{eqnarray}

The contribution is split into the first, energy-conserving term, and the second term, oscillating with time, in
which energy is not conserved at the vertex $1$ %this is a label
 \cite{Dadic:2009}.

The tadpole counterterm follows the same pattern:

\begin{eqnarray}\label{gtadcount}
&&
G^{tadpole}_{count,j}(x_2)
=-G_{A}(0,0)
%\cr\nonumber\\&&
 +\int{dp_{02}\over 2\pi}\, {ie^{ip_{02}x_{02}}\over p_{02}-i\epsilon}
\, [G_{A}(p_{02},0)-G_{A}(0,0)].
\end{eqnarray}

Notice the similarity of the expressions (\ref{gtad}) and (\ref{gtadcount}).

An important point here is that the tadpole contribution splits into two:
 (1) the energy-conserving part and (2) the energy nonconserving part.

 In the energy conserving part, the constant multiplying the counterterm
 may be adjusted to satisfy the
renormalization condition $\langle 0|\phi|0 \rangle = 0$ of the S-matrix theory,
by which the tadpoles are completely eliminated from
 perturbation expansion. Nevertheless, the terms proportional to $f$ survive.
 The energy nonconserving terms oscillate with time, with the frequency depending on the energy increment. In the competition with the contributions of subdiagram without tadpoles, they fade with time,
 thus giving the same
 $t\rightarrow \infty $ limit as expected from S-matrix theory.

The $g^3$ order tadpoles and tadpoles with the resummed loop propagator
(obtainable after renormalizing the self-energy; see further in the text) do not change our conclusions.

\subsection{UV Divergence at the Self-Energy Subdiagram}

While in the S-matrix theory, there is only Feynman
 ($\Sigma^1_{F}(p_0,\vec p)$) one-loop self energy, which does not
 depend on the frame, in out-of-equilibrium FT, we have self-energies
 $\Sigma^1_{R}(p_0,\vec p)$, $\Sigma^1_{A}(p_0,\vec p)$, and $\Sigma^1_{K}(p_0,\vec p)$, which is frame dependent through $f(\omega_p)$ (notice here that we distinguish the ``true'' retarded and advanced
  functions from those that carry index $R$ ($A$), but do not vanish for
  $t_2>t_1$ ($t_2<t_1$), except at $d<4$).
%
%\begin{eqnarray}\label{g2RR}
%&&
%\Sigma^1_{F}(p_0,\vec p) = i^2\, g^2\, \mu^{\kappa}\,\int {d^dq_1 d^dq_2\over (2\pi)^{2d}}G_{R,f=0}(q_1)G_{R,f=0}(q_2)
%(2\pi)^d\delta^{(d)}(q_1+q_2-p_1-p_2).
%\end{eqnarray}
% 
\begin{eqnarray}\label{RAK1}
&&
\Sigma^1_{R}(p_0,\vec p)=-ig^2\mu^{\kappa} \int {d^dq\over 2(2\pi)^{d}} 
[G_{R}(p_0-q_0,\vec p-\vec q)G_{K,R}(q_0,\vec q)
\cr\nonumber\\&&
+G_{K,R}(p_0-q_0,\vec p-\vec q)G_{R}(q_0,\vec q)]
=\Sigma^{1,*}_{A}(p_0,\vec p),
\cr\nonumber\\&&
\Sigma^1_{K}(p_0,\vec p)
=-\Sigma^1_{K,R}(p_0,\vec p)+\Sigma^1_{K,A}(p_0,\vec p)
\end{eqnarray}
\begin{eqnarray}\label{RAK1k}
&&
\Sigma^1_{K,R}(p_0,\vec p)=
-i\, g^2\, \mu^{\kappa} \int {d^dq\over 2(2\pi)^{d}} 
\, [G_{K,R}(p_0-q_0,\vec p-\vec q)G_{K,R}(q_0,\vec q)
\cr\nonumber\\&&
+ \, G_{R}(p_0-q_0,\vec p-\vec q) q, G_{R}(q_0,\vec q)]
=-\Sigma^{1,*}_{K,A}(p_0,\vec p).
\end{eqnarray}

Now, all the integrals containing $f(\omega_p)$ are UV finite owing to the assumed UV cut-off in the definition of $f$.
 Vacuum contributions to $\Sigma^1_{K,R}$ are
 finite separately at $d\rightarrow 4$; at
  $d\rightarrow 6$, this is no longer the case, but their sum is finite.

%\newpage
For retarded and advanced self-energies, imaginary parts and parts
 proportional to $f(\omega_p)$ are UV finite and could be
 calculated directly from (\ref{RAK1}).
Real, vacuum parts of $\Sigma^1_{R}$ are connected to
 $\Sigma^1_{F}$, and we use the results already available
 from S-matrix renormalization.
The connection is:

\begin{eqnarray}\label{g2KRR}
&&
\Sigma^1_{j,k}={1\over 2}[-\Sigma^1_{K,R}+\Sigma^1_{K,A}
-(-1)^k\Sigma^1_{R}-(-1)^j\Sigma^1_{A}],
\cr\nonumber\\&&
Re\Sigma^1_{R,f=0}=Re\Sigma^1_{A,f=0},~~
 Re\Sigma^1_{K,R,f=0}=Re\Sigma^1_{K,A,f=0},
 \cr\nonumber\\&&
Re\Sigma^1_{F}=Re\Sigma^1_{R,f=0}.
\end{eqnarray}

%
%ADD
{The regularization procedure (either by making $d<4$ or by
 introducing fictive massive particles as in Pauli--Villars regularization)
 is usually considered artificial. Nevertheless,
 there are efforts to generate necessary massive particles 
 (virtual wormholes) dynamically \cite{kirilov:2015}. }
 %ADDEND

For $ \Sigma^1_{F}(p) $, we find in the literature {\cite{RyderBook}}:
\begin{eqnarray}\label{g2F}
&&
\Sigma^1_{F}(p)={1\over 2}i^2g^2{\int d^4q_1 d^4q_2
\over (2\pi)^8}G_F(q_1)G_F(q_2)
(2\pi)^4\delta^{(4)}(q_1-q_2-p),
\cr\nonumber\\&&
={1\over 2}g^2{\int d^4q_1 d^4q_2\over (2\pi)^8}
{(2\pi)^4\delta^{(4)}(q_1-q_2-p)\over (q_1^2-m^2+i\epsilon)
(q_2^2-m^2+i\epsilon)},
\cr\nonumber\\&&
\Longrightarrow 
{1\over 2}g^2(\mu)^{\kappa}\int^1_0dz\int {d^dq'\over (2\pi)^d}
{1\over [q'^2-m^2+p^2z(1-z)+i\epsilon]^2},
\cr\nonumber\\&&
={ig^2\over 32\pi^2}(\mu^2)^{\kappa/2}\Gamma(\kappa/2)
\int^1_0dz[{p^2z(1-z)-m^2+i\epsilon \over 4\pi\mu^2}]^{-\kappa/2}.
\end{eqnarray}

The last relation above is still causal. It is UV finite, and it allows the 
separation into the sum of the retarded and advanced term. However, the expansion of
 $[{p^2z(1-z)-m^2+i\epsilon / 4\pi\mu^2}]^{-\kappa/2}$
 in power series of $|\kappa|$ is allowed only when
 $\kappa\ln[p^2/(4\pi\mu)]<<1$; thus, it is a ``low energy'' expansion,
 and in spite of the fact that $\kappa $ may be taken arbitrarily small,
 the limit $|p_0|\rightarrow \infty $ is never allowed.

\begin{eqnarray}\label{g2Fa}
&&
\Sigma^1_{F}(p)
\approx {ig^2\mu^{\kappa}\over 16\pi^2(\kappa)}-{ig^2\mu^{\kappa}\over 32\pi^2}
[\, \gamma_{_E} + \int^1_0dz\ln[{p^2z(1-z)-m^2+i\epsilon\over 4\pi\mu^2}] \, ]
\cr\nonumber\\&&
= {ig^2\mu^{\kappa}\over 16\pi^2(\kappa)}+finite.
\end{eqnarray}

This expression is no longer causal; it is valid only if
$\kappa\ln[p^2/(4\pi\mu)]<<1$.
 One needs the vanishing of
 self-energy for $|p_0| \rightarrow \infty $, i.e., the region where the opposite
 condition
 $\kappa\ln[p^2/(4\pi\mu)]>>1$ is fulfilled. Then, $|\Sigma^1_{\infty,F}(p)|\rightarrow 0$ as $|p_0| \rightarrow \infty $ as far as $\kappa\neq 0 $.

 The integration over $z$ gives:
 \begin{eqnarray} \label{2detailedSelf-energy}
&&
\Sigma^1_{F}(p) = - \frac{g^2}{16 \pi^2} \left\{\frac{1}{\kappa} -\frac{\gamma_{_E}}{2} + 1
 +\frac{1}{2} \ln(4\pi \frac{\mu^2}{m^2}) - \frac{1}{2} \sqrt{1-\frac{4 m^2}{p^2+i\epsilon}}
%\cr\nonumber\\&&
\ln\!\left[\frac{\sqrt{1-\frac{4 m^2}{p^2+i\epsilon}} + 1}{\sqrt{1-\frac{4 m^2}{p^2+i\epsilon}}- 1} \right] 
\right\}
\end{eqnarray}
with a high $p_0$ limit:
\begin{eqnarray} \label{2detailedSelf-energylarge}
&&\!\!\!\!\!\Sigma_F(p^2,m^2)_{p^2\rightarrow \infty} \approx -\frac{g^2}{16 \pi^2} \left\{\frac{1}{\kappa}
 -\frac{\gamma_{_E}}{2} + 1 +\frac{1}{2} \ln(4\pi \frac{\mu^2}{m^2}) - \frac{1}{2} 
%\cr\nonumber\\&&
\ln\!\left[-\frac{ m^2}{p^2} \right] 
\right\}.
\end{eqnarray}

To verify the causality of the two-point function, one may try to project out the
retarded part of the finite (subtracted) part of $ \Sigma^1_{F}(p) $, namely
$-i \int {dp'_0 \over 2\pi}\, \Sigma^1_{F, finite}(p) / (p_0-p'_0-i\epsilon)$,
by integration $\int dp_0$ over a large semicircle. However, the contribution over
 a very large semicircle does not vanish, and the integral is ill defined.

 Indeed, we have started from the expressions for $G_F$ ($\Sigma_F $)
 containing only
retarded and advanced functions, and in the absence of divergence, we
expect this to be the truth at the end of calculation. Instead, the function
in the last two lines of Expression
(\ref{g2Fa}) is not a combination of the $R$ and $A$ functions, otherwise it
 should vanish when $|p_0|\rightarrow\infty$ and $\kappa $
are chosen as arbitrarily small; such a behavior can be shifted to
an arbitrarily high scale. However, the limit $\kappa\rightarrow 0$ remains
always out of reach.
To preserve causality, we should keep the whole $p_0$ complex plane. Specifically,
we need the region with large $|p_0|$, to be able to
 integrate over a large semicircle in the complex $p_0$ plane, at least to get
 $\int dp_0\Sigma^1_{R}(p)G_{K,A}(p_0)=0$.
Thus, we have obtained a result correct at $\kappa\neq 0$ and problematic
at $d=4$.

Fortunately enough, there is a way to ``repair'' causality: the composite object
$G_{F}(p)\Sigma^1_{F}(p)$ is vanishing when $|p_0|\rightarrow\infty$;
 it can be split into its retarded and advanced parts; thus, it is causal.
 This sort of reparation of causality is possible in other QFT in which
 logarithmic UV divergence appears. It is similar to the Glaser--Epstein
 \cite{Epstein:1973gw,G. Scharf:1995,Gracia:2003} approach, where not just $\Sigma $, but
$G\Sigma $ are the subjects of expansion.

In this spirit, we agree with the conclusion of \cite{Millington:2013,Millington:2013pr,Millington:2014}:
``Our amplitudes are manifestly
causal, by which we mean that the source and detector are
always linked by a connected
chain of retarded~propagators.''

\begin{figure}[H]

\centering

\includegraphics[width=12 cm]{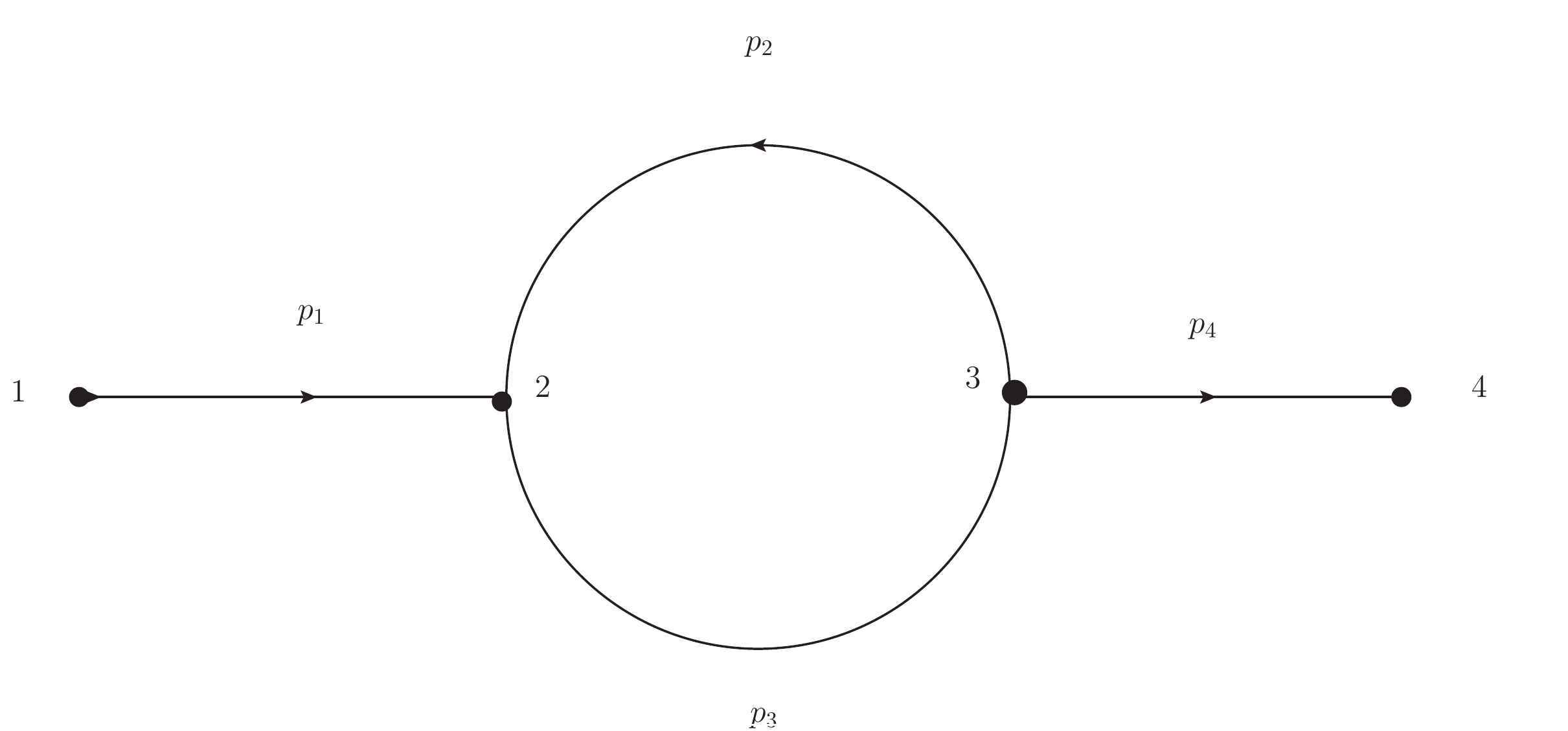}

\caption{The vertex diagram.\label{fig2}}

\end{figure}

Similar is the problem we can see by considering
$\lambda \phi^4$ theory. In this theory, the loop of Figure \ref{fig2} is a vertex diagram,
and the above Glaser--Epstein philosophy does not apply. Nevertheless, the propagator attached to the vertex depends on $p_0$ and ``improves'' the convergence of $dp_0$ integration.

\subsection{Self-Energy Diagram with Legs}

To be able to introduce composite objects with $\Sigma_{R(A)}$, we need one of $\Sigma_{R(A)}$'s vertices to conserve energy. The lower in time vertex
 may be the minimal time vertex, so it does not help in all cases. However, the
 higher in time vertex would do it, if both the integrals $dq_0$
 and $dp_0$ converge.

The $\Sigma_{ij}$ self-energy contributions with legs (see Figure \ref{fig2}) are:
\begin{eqnarray}\label{gse}
&&
G_{R}\Sigma^1_{K,R}*G_{A},~
%\cr\nonumber\\&&
G_{R}*\Sigma^1_{K,A}G_{A},~
%r\nonumber\\&&
G_{R}\Sigma^1_{R}*G_{K,A},~
%\cr\nonumber\\&&
G_{K,R}*\Sigma^1_{A}G_{A},
\cr\nonumber\\&&
G_{R}\Sigma^1_{R}*G_{A},~
%\cr\nonumber\\&&
G_{A}\Sigma^1_{A}G_{A} ,~
%\cr\nonumber\\&&
G_{R}\Sigma^1_{R}G_{K,R},~
%\cr\nonumber\\&&
G_{K,A}\Sigma^1_{A}G_{A}.
\end{eqnarray}
%}

In the above expression, $\Sigma$s are introduced {in Equation} (\ref{RAK1}).
``$*$'' indicates the convolution product, which includes the energy nonconserving vertex.
Terms containing $\Sigma^1_{K,R}$ and $\Sigma^1_{K,A}$ are UV-finite, creating no problems. The other terms, containing
$\Sigma^1_{R}$ and $\Sigma^1_{A}$, are finite as long as
$d<4$, and we may obtain their real part through (\ref{g2F}).

Two features seem potentially suspicious:
(1) UV divergence in the loop defining $\Sigma^1_{R(A)}$,
(2) the ill-defined vertex function between $G_{R}$ and $\Sigma^1_{R}$
and between $\Sigma^1_{A}$ and $G_{A}$.

 Nevertheless, both problems are resolved at $d<4$: ``to be'' UV divergence is
 subtracted and energy conservation is recovered in the above-mentioned vertices.
 The composite objects $G_{R}(p)\Sigma^1_{R}(p)$ and
 $\Sigma^1_{A}(p)G_{FA}(p)$ are now well defined.

%%%%%%%%%%%%%%%%%%%%%%%%%%%%%%%%%%%%%%%%%%

\section{Discussion and Conclusions}

We examined renormalization prescriptions for the finite-time-path out-of-equilibrium
 $\lambda \phi^3$ QFT in the basis of $G_R, G_A, G_{K,R}$, and $G_{K,A}$ propagators.

As expected, the number of
counterterms did not change, and
the formalism enables term by term finite perturbation calculation.

There are some interesting features:

\begin{enumerate}[leftmargin=*,labelsep=4.9mm]

\item The integrals ensuring the energy conservation at the
vertices above $\Sigma_R$ and $\Sigma_A$ should have been done before
taking the limit $d=4$.

\item The renormalized self-energies ($\Sigma_{F}$, $\Sigma_{R}$, and $\Sigma_{A}$) are not a linear combination of true retarded and advanced
 components. This is directly readable from the
  final result, which does not
 vanish as $|p_0|\rightarrow \infty$ in all directions in a complex plane $p_0$.
 This problem is present already in S-matrix theory, and we only recognize
 it properly as a causality problem, in the sense
 that the expected properties of the theta-function fail:
 $\Theta(t)\Theta(-t)\neq 0$ or $\Theta(t)\Theta(t)\neq \Theta(t)$.
 While it is not clear what harm it does to the theory, one may introduce
 ``composite objects'' $G_{F}(p)\Sigma^1_{F}(p)$,
 $G_{R}(p)\Sigma^1_{R}(p)$, and
 $\Sigma^1_{A}(p)G_{A}(p)$ {to} improve convergence,
 and the causality is ``repaired''. Indeed in the
 Glaser--Epstein approach, they consider the perturbation expansion, in which
 only self-energy with a leg appears.

\item The tadpole contribution splits into the energy-conserving, constant component,
which is eliminated by renormalization condition, and the other energy nonconserving, time-dependent component, is finite after subtraction.
These tadpole contributions are strongly oscillating with time and vanish as $t\rightarrow \infty $,
 in good agreement with the renormalization condition $<0|\phi|0>=0$ of the S-matrix~theory.

\item The regularization ($d\neq 4$) is extended till the late phase of calculation.
\end{enumerate}

 The procedure is therefore generalized for application to more realistic theories (QED and QCD, electro-weak QFT, etc.) by the following:

 (A) regularize; (B) do energy-conserving integrals; (C) subtract ``to be'' UV infinities;
 (D) deregularize (do limit $d\rightarrow 4$).

 Again, the above described Features (1) and (2) will emerge.

 This work contains many of the features \cite{Urmossy:2016} arising in the more realistic
theories like QED or QCD. Such finite-time-path renormalization
 is a necessary prerequisite for the calculation of damping rates,
 and other transition coefficients under the more realistic conditions truly away
 from equilibrium as opposed to the results obtained within the linear response
 approximation.

 Our plan is to extend the exposed methods to the case of QED. Specifically,
 we resolve the controversy of the UV diverging
 number of direct photons in the lowest order of quark QED,
 as calculated by Boyanovsky and {collaborators}
 \cite{Boyanovsky:2001,Boyanovsky:2002}
 and criticized by \cite{Arleo:2004gn}.
 We find that, at the considered one-loop
 order of perturbation, it is only the vacuum-polarization diagram contributing.
 The renormalization leaves only finite contributions to the
 photon production \cite{dadickk}.

\authorcontributions{Conceptualization,  I.D. and D.K.; Formal analysis,  I.D.;
Investigation,  I.D. and D.K.; Methodology,  I.D. and D.K.;  Validation,  I.D. and D.K.;
Visualization,  D.K.; Writing – original draft,  I.D.; Writing – review \& editing,  I.D. and D.K.}
%%%%%%%%%%%%%%%%%%%%%%%%%%%%%%%%%%%%%%%%%%
%For research articles with several authors, a short paragraph specifying their individual contributions must be provided. The following statements should be used “Conceptualization, X.X. and Y.Y.; Methodology, X.X.; Software, X.X.; Validation, X.X., Y.Y. and Z.Z.; Formal Analysis, X.X.; Investigation, X.X.; Resources, X.X.; Data Curation, X.X.; Writing-Original Draft Preparation, X.X.; Writing-Review & Editing, X.X.; Visualization, X.X.; Supervision, X.X.; Project Administration, X.X.; Funding Acquisition, Y.Y.”, please turn to the CRediT taxonomy for the term explanation. Authorship must be limited to those who have contributed substantially to the work reported.

\funding{This work was supported in part by the Croatian Science Foundation under
 Project Number 8799 and by STSM%define if appropriate
 grants from COST Actions CA15213 THOR%define if appropriate
 and
 CA16214 PHAROS%define if appropriate
.}

\conflictsofinterest{The authors declare no conflict of interest.} 
%%%%%%%%%%%%%%%%%%%%%%%%%%%%%%%%%%%%%%%%%%
%%% Please add these part.

%%%%%%%%%%%%%%%%%%%%%%%%%%%%%%%%%%%%%%%%%%

\abbreviations{The following abbreviations are used in this manuscript:\\

\noindent 
\begin{tabular}{@{}ll}

QFT & quantum field theory\\

FTP & finite-time-path \\

RT & renormalization theory \\

DR & dimensional regularization \\

UV & ultra-violet \\         

QED & quantum electrodynamics \\

QCD & quantum chromodynamics \\

\end{tabular}}

%%%%%%%%%%%%%%%%%%%%%%%%%%%%%%%%%%%%%%%%%%

\appendixtitles{yes} %Leave argument "no" if all appendix headings stay EMPTY (then no dot is printed after "Appendix A"). If the appendix sections contain a heading then change the argument to "yes".
\appendixsections{yes}
\appendix
\section{}

This Appendix provides the derivation of Equation (\ref{gtad}).

The tadpole diagram, Figure \ref{fig1}, appears as a propagator with both ends attached to the same vertex.
 We start in coordinate representation.
To sum contributions from the vertices of Types $1$ and $2$, we write the propagators with the help
of the Wigner transform. Keldysh-time-path propagators and the finite-time propagators become
 identical in the limit $t' \to \infty $.
To translate to the $R/A$ basis, we use $G_{i,j}={1\over 2}[G_K+(-1)^jG_R+(-1)^{i}G_A]$.

\begin{eqnarray}\label{gtadDerive}
&&
G_{tad,j}(x_2)=ig\mu^{\kappa/2}\int d^dx_1
\cr\nonumber\\&&
\times[G_{1,1} (x_1,x_1)G_{1,j}(x_1,x_2)-G_{2,2} (x_1,x_1)G_{2,j}(x_1,x_2)],
\cr\nonumber\\&&
=ig\mu^{\kappa/2}\int d^{d-1}x_1\int_0^{\infty} dx_{01}e^{-ip_{2}(x_{1}-x_{2})}
{d^{d}p_1\over (2\pi)^d}{d^{d}p_2\over (2\pi)^d}
\cr\nonumber\\&&
\times[G_{1,1,x_{01}} (p_{1})G_{1,j,t}(p_{2})-G_{2,2,x_{01}} (p_{1})G_{2,j,t}(p_{2})],~~t={x_{01}+x_{02}\over 2},
\cr\nonumber\\&&
=ig\mu^{\kappa/2}\int d^{d-1}x_1\int_0^{\infty} dx_{01}{d^{d}p_1\over (2\pi)^d}{d^{d}p_2\over (2\pi)^d}
\cr\nonumber\\&&
\times e^{-ip_{2}(x_{1}-x_{2})}dp'_{01}dp'_{02}P_{x_{01}}(p_{01},p'_{01})P_{t}(p_{02},p'_{02})
\cr\nonumber\\&&
\times[G_{1,1,\infty} (p'_{1})G_{1,j,\infty}(p'_{2})-G_{2,2,\infty} (p'_{1})G_{2,j,\infty}(p'_{2})],
\cr\nonumber\\&&
p'_1=(p'_{01},\vec p_1),~p'_2=(p'_{02},\vec p_2),
\end{eqnarray}
where we have used the projection operator $P$ connecting time-dependent lowest order propagators
 with time-independent lowest order propagators \cite{Dadic:2001bp,erratumDadic:2001bp}:

\begin{eqnarray}\label{RAK0}
&&G_{t}(p_0,\vec p)=\int_{-\infty}^{\infty}dp'_0 
P_t(p_0,p_0')G_{\infty}(p_0',\vec p),
\cr\nonumber\\&&
P_t(p_0,p_0')= {\Theta(t) \over 2 \pi}\int_{-2t}^{2t}ds_0e^{is_0(p_0-p_0')} = 
{\Theta(t) \over \pi} {\sin 2(p_0-p_0')t \over (p_0-p_0')},
\cr\nonumber\\&&
\lim_{t \rightarrow \infty}P_t(p_0,p_0')=\delta(p_0-p_0'),
\cr\nonumber\\&&
\int_{-\infty}^{\infty}dp_0 e^{-is_0p_0}P_t(p_0,p'_0)=e^{-is_0p'_0}\Theta(t)\Theta(2t-s_0)\Theta(2t+s_0).
\end{eqnarray}

Here, $G$ is a bare propagator (matrix propagator or $R$, $A$, or $K$ propagator.)

A similar relation holds for lowest order self-energies:
\begin{eqnarray}\label{sigma1}
\Sigma^1_{t}(p_0,\vec p)=\int_{-\infty}^{\infty}dp'_0 P_t(p_0,p_0')
\Sigma^1_{\infty}(p_0',\vec p),
\end{eqnarray}
where $\Sigma^1_{t}$ is the retarded, advanced, or Keldysh self-energy.

By using the above relations, we obtain:

\begin{eqnarray}\label{gtadDerive2}
&&
G_{tad,j}(x_2)
=ig\mu^{\kappa/2}\int d^{d-1}x_1\int_0^{\infty} dx_{01}e^{-ip'_{2}(x_{1}-x_{2})}
{d^{d}p'_1\over (2\pi)^d}{d^{d}p'_2\over (2\pi)^d}
\cr\nonumber\\&&
\times[G_{1,1,\infty} (p'_{1})G_{1,j,\infty}(p'_{2})-G_{2,2,\infty} (p'_{1})G_{2,j,\infty}(p'_{2})],
\cr\nonumber\\&&
=ig\mu^{\kappa/2}(2\pi)^{-1}\int{-i\over p'_{02}-i\epsilon}\delta^{(d-1)}(\vec p'_2)e^{ip'_{02}x_{02}}{d^{d}p'_1\over (2\pi)^d}d^dp'_{2}
\cr\nonumber\\&&
\times[G_{1,1,\infty} (p'_{1})G_{1,j,\infty}(p'_{2})-G_{2,2,\infty} (p'_{1})
G_{2,j,\infty}(p'_{2})],
\cr\nonumber\\&&
=ig\mu^{\kappa/2}(2\pi)^{-1}\int{-i\over p'_{02}-i\epsilon}
\delta^{(d-1)}(\vec p'_2)e^{ip'_{02}x_{02}}{d^{d}p'_1\over (2\pi)^d}d^dp'_{2}
\cr\nonumber\\&&
\times{1\over 2}[-G_{K,\infty} (p'_{1})G_{A,\infty}(p'_{2})
-G_{R,\infty} (p'_{1})G_{K,\infty}(p'_{2})
-G_{A,\infty} (p'_{1})G_{K,\infty}(p'_{2})
\cr\nonumber\\&&
+(-1)^jG_{R,\infty} (p'_{1})G_{R,\infty}(p'_{2})
+(-1)^jG_{A,\infty} (p'_{1})G_{R,\infty}(p'_{2})],
\end{eqnarray}

By taking the fact that tadpoles with $G_R$ and $G_A$ vanish, we obtain:

\begin{eqnarray}\label{gtadDerive3}
&&
G_{tad,j}(x_2)
=ig\mu^{\kappa/2}{(2\pi)^{-1}\over 2}\int{i\over p'_{02}-i\epsilon}
\delta^{(d-1)}(\vec p'_2)e^{ip'_{02}x_{02}}
\cr\nonumber\\&&
\times{d^{d}p'_1\over (2\pi)^d}d^dp'_{2}
G_{K,\infty} (p'_{1})G_{A,\infty}(p'_{2}),
\cr\nonumber\\&&
=(2\pi)^{-1}\int{i\over p'_{02}-i\epsilon}
e^{ip'_{02}x_{02}}G_{A,\infty}(p'_{02},0)dp'_{02}G_{Tad}
\cr\nonumber\\&&
G_{Tad}={ig\mu^{\kappa/2}\over 2}\int G_{K,\infty} (p'_{1}){d^{d}p'_1\over (2\pi)^d}.
\end{eqnarray}

Thus,
\begin{eqnarray}
G_{tad,j}(x_2)
=-G_{A,\infty}(0,0)
G_{Tad}
%\cr\nonumber\\&&
+\int{dp'_{02}\over 2\pi}{ie^{ip'_{02}x_{02}}\over p'_{02}-i\epsilon}
[G_{A,\infty}(p'_{02},0)-G_{A,\infty}(0,0)]G_{Tad}.
\end{eqnarray}

The contribution is split into the first, energy-conserving term, and the second term, oscillating with time, in
which energy is not conserved at the vertex $1$%this a label
.

%%%%%%%%%%%%%%%%%%%%%%%%%%%%%%%%%%%%%%%%%%

\reftitle{References}

%%%%%%%%%%%%%%%%%%%%%%%%%%%%%%%%%%%%%%%%%%

\end{document}